\documentclass[runningheads]{llncs}
\usepackage[T1]{fontenc}
\usepackage{graphicx}

\usepackage{hyperref}
\begin{document}
\title{Gaze2AOI: Open Source Deep-learning Based System for Automatic Area of Interest Annotation with Eye Tracking Data}
\titlerunning{Gaze2AOI}
%

\author{Karolina Trajkovska\inst{1} \and
Matjaž Kljun\inst{1,2} \and
Klen Čopič Pucihar\inst{1,2,3}}

\authorrunning{K. Trajkovska et al.}
%
\institute{Faculty of Mathematics, Natural Sciences and Information Technologies, University of Primorska, Koper, Slovenia, 
\and Department of Information Science, Stellenbosch University, South Africa,
\and Faculty of Information Studies, Novo Mesto, Slovenia.
}

%
\maketitle              
\begin{abstract}
Eye gaze is considered an important indicator for understanding and predicting user behaviour, as well as directing their attention across various domains including advertisement design, human-computer interaction and film viewing. In this paper, we present a novel method to enhance the analysis of user behaviour and attention by (i) augmenting video streams with automatically annotating and labelling areas of interest (AOIs), and (ii) integrating AOIs with collected eye gaze and fixation data. The tool provides key features such as time to first fixation, dwell time, and frequency of AOI revisits. By incorporating the YOLOv8 object tracking algorithm, the tool supports over 600 different object classes, providing a comprehensive set for a variety of video streams. This tool will be made available as open-source software, thereby contributing to broader research and development efforts in the field.


\end{abstract}

\begin{figure}
\centering
    \includegraphics[width=\textwidth]{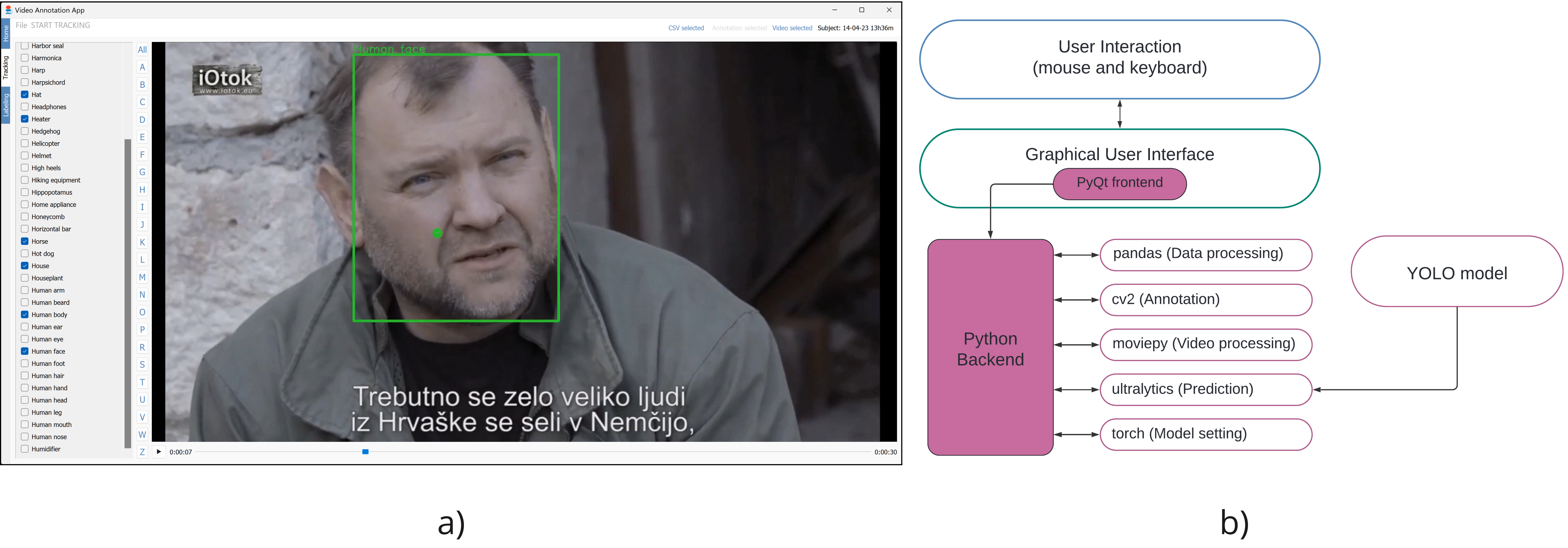}
    \caption{Gaze2AOI: (a) User interface for selecting object classes and performing object tracking; (b) System architecture.
    }
    \label{fig:teaser}
\end{figure}

\section{Introduction}

One of the challenges in video consumption research is understanding and predicting users' attention and behaviour. The models designed to predict user behaviour can be based on a range of predictor variables captured from various devices and tools. Technological advancements in physiological sensing  have greatly enhanced and simplified the process of capturing user data. Nowadays, we can reliably perform eye and facial expression tracking, heart rate monitoring, measuring the electrical characteristics of the skin, and even monitoring electrical brain activity, to gain a deeper understanding of user behaviour.


For example, research has demonstrated the potential to link eye movement with the brain and neurological processes, providing valuable insights into user's mental state ~\cite{bell1823xv,wadeMoving,carter2020best}. 
Beyond analysing physiological responses of the eye, such as pupil size, blink rate, gaze velocity and frequency, researchers have also recognised the importance of understanding the semantic meaning behind eye gaze. Knowing what users look at and when is a key indicator of how visual attention is allocated~\cite{carter2020best} and what users' interests are. 
This greatly enhances our potential to develop systems that adapt to and model users' attention. 

Understanding what users look at is traditionally achieved by using Areas of Interest (AOI), which denote specific regions within a watched scene with semantic significance, such as a particular object or event within a scene~\cite{blascheck2014state}. By coupling AOI with other metrics, such as transition counts (i.e., movement from one AOI to another), AOI revisit frequency, dwell duration within an AOI, and the time to the first fixation (TTFF), we can gain valuable insights into user behaviour. However, acquisition of data from multiple physiological sensors poses a challenge of how to effectively clean, transform and aggregate this data. In addition, semantic analysis of eye tracking data can also prove difficult when dealing with dynamic unpredictable scenarios, such as when we are using eye tracking glasses (e.g. difficult to predict where users will look at), or when the content within screen eye trackers is dynamic (e.g. the semantic information about what is shown on the screen is not known).

Despite significant advances in video segmentation and object detection~\cite{xiao2020review}, the segmentation and labelling of AOI combined with eye tracking data are predominantly performed manually--a labour-intensive and time-consuming task. To the best of our knowledge, no open-source tool currently exists that automates this process, which is the focus and the main contribution of this paper.   
We introduce a novel open-source tool that leverages on a deep learning model based on the YOLO algorithm, capable of distinguishing around 600 object classes to automatically detect and label AOIs. Besides automatically labelling AOIs, the tool can generate standard AOI metrics, such as transition counts, number of revisits, dwell duration, and time to the first fixation (TTFF). Additionally, the tool offers customised labelling options to further enhance AOI semantics.

\section{Related Work}

There have been several attempts to employ computer vision approaches for automatic AOI generation. These include using feature descriptors, such as  \textit{scale invariant feature transform} (SIFT)~\cite{beugher} and \textit{Canny Edge Detector}~\cite{fehringer}. The main problem with these systems is a lack of flexibility and poor performance when large number of objects need to be detected and tracked in a strongly dynamic environment. 
According to Wolf et al.~\cite{wolf2018automating} and Duchowski et al.~\cite{duchowski}, the prevailing solutions used until 2018 were manual gaze mapping and marker-based techniques. However, it has been stressed that these solutions are not suited for eye tracking studies with lengthy recording durations. One of the breakthroughs was provided by a deep learning architecture called region-based convolutional neural network (R-CNN), which facilitates the automated mapping of gaze data onto corresponding AOIs~\cite{wolf2018automating}. The model has been integrated into an open-source tool, but it requires large training data sets in a form of manually annotated images. 

Zhang et al.~\cite{zhang2018complete} presented another tool aimed at streamlining the entire process from collecting eye tracking data to its analysis and annotation. The tool offers automatic AOI detection combined with manual labelling by the user. However, the automation is limited to human faces and is achieved using Haar-based face detection. One of the latest automated identification of nonhuman AOI is the web-based tool eyeNotate~\cite{barz}. It is based on image classification on a $200 \times 200$ pixel patch that has been cut around the fixation point of the video frame~\cite{barz2021automatic}. The tool provides various features, including the option to adjust the accuracy threshold for automatic suggestions 
and flexibility to provide own labels. However, in many research contexts, the significance lies in what users did not observe, such as a specific trademark in an advertisement. In such cases, the AOI would not be engaged, thus escaping detection by the algorithm that generates a patch around the fixation point. 
Moreover, the tool eyeNotate is not open source, but rather accessible to the public as a web application and is currently not accessible.






In this paper we present and describe the Gaze2AOI tool and offer it to the community as an open-source software. The tool was initially developed to analyse eye tracking behaviour in an interactive documentary~\cite{julie,ducasse2022interactive}. It includes a pre-trained model based on the YOLOv8 algorithm capable of recognising and annotating around 600 distinct human and non human object classes across a variety of domains, as well as objects proximal to the fixation point or those inadvertently overlooked (\autoref{fig:teaser} (a)). This model, sourced from Ultralytics~\cite{ultralyticsHome} and pre-trained on October 2022 release of the OpenImagesv7 dataset~\cite{storageOpenImages,OpenImages}, offers versatility by accommodating various datasets on which YOLO pre-trained models are available or can be trained. Our tool is designed for seamless adaptation to any model from Ultralytics, fostering its applicability across different research endeavours.


\section{Gaze2AOI}

The Gaze2AOI tool features 
the ability to perform AOI annotations, and, if desired, customised labelling for each key-frame where key-frames are defined as frames where new a new object is detected. It supports videos of any frame rate and of varying lengths. Notably, the system operates autonomously without requiring an internet connection, ensuring seamless usability across different scenarios. In the following section, an in-depth explanation of the system architecture, incorporated technologies, key functionalities, and user interface are provided.




\subsection{User Interface}

The landing page includes instructional content. It can be accessed at any time by clicking on the tab labelled ``Home'' in a vertical menu bar on the right side of the window. Two additional pages--one dedicated to run object tracking (i.e. AOI) predictions, and one for optional labelling--can be accessed via tabs labelled ``Tracking'' and ``Labelling''. Users can move between tabs without losing current data on other tabs (see \autoref{fig:teaser} (a)). On the object tracking and labelling pages, an additional horizontal menu bar allows file selection (\autoref{fig:teaser} (a)). On the right side of this horizontal bar, a label for each file to be selected lights up when the file is selected. All selected files should pertain to the same subject. If this is not the case, a red note will appear after they are selected.

The page for performing object tracking features two panels (\autoref{fig:teaser} (a)). The left panel displays all object classes as a list of check-boxes arranged alphabetically. Additionally, users can filter the list by the first letter of the object class for ease of navigation. After selecting the files to be processed and the desired check-boxes for classes, users can initiate the tracking process by clicking the ``Start tracking'' button. Once tracking is complete, users can preview the annotated video in the main panel on the right showing a visual representation of the tracked objects, and review the outcomes of the tracking process. Additionally, the annotation of the fixation point is included in the preview (see small purple dot (\autoref{fig:teaser} (a))).
To provide a clear visual feedback on the relationship between the fixation points and identified objects (i.e., the AOIs) the objects are colour-coded: green if the fixation point falls within the bounding box of the AOI (\autoref{fig:labeling}), or red otherwise.  

As mentioned, the system provides an option to add additional labels to an AOI, in addition to the automatically assigned label. 
This feature is available on the ``Labelling'' page (\autoref{fig:labeling}), which can be accessed from the side menu tabs. On this page users encounter a split-panel interface: the left panel displays the annotated frame, while the right panel presents the names of the detected objects alongside an input field where customised labels can be entered. The names of the detected objects are also colour-coded to provide visual cues: the names of AOIs associated with fixation points are green, while the names of AOIs that were overlooked are red. Users have the flexibility to label both types of AOIs according to their specific requirements to refine semantic meaning of AOIs.  For example, users can provide names of people there were identified as a human face.

\subsection{Object Detection and Tracking}

``You Only Look Once'' (YOLO)~\cite{redmon2016you} is a deep learning algorithm used for object tracking. The Ultralytics YOLO~\cite{ultralyticsHome} model offers multiple advantages, 
such as (i) a choice between two distinct trackers namely BoT-SORT and ByteTrack, with the latter being used in the current implementation, and (ii) the flexibility that allows users to easily modify the model version, settings, or the selected tracker within the open-source tool we have developed. This adaptability ensures that users can tailor the tracking approach to suit their specific requirements. 

The predictions are acquired for every frame within the video. The outcomes are associated with the eye-tracking data based on the frame number. The output result for each identified object in a frame is the object name and a tuple of three values that indicates whether the item is detected, gazed upon or contains the fixation point on that frame. The presence of this particular output facilitates the process of data aggregation and the acquisition of metrics such as the time to first fixation (TTFF), the revisits of an AOI, and dwell duration within AOI. 


\subsection{Customised Labelling}

\begin{figure}
\centering
\includegraphics[width=\textwidth]{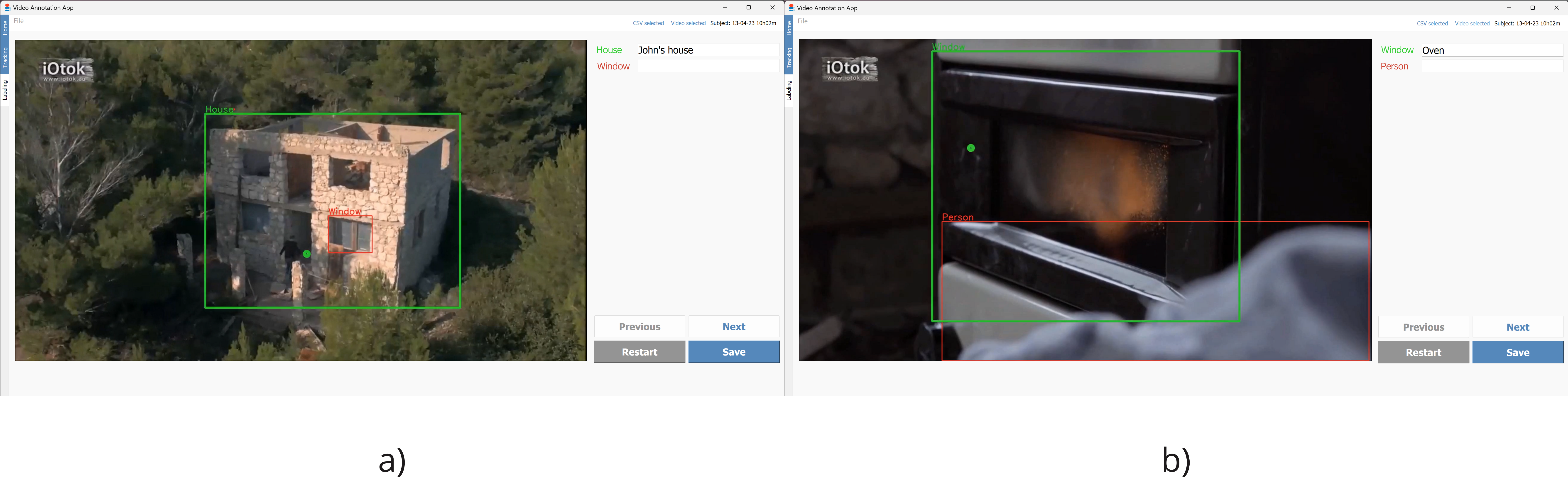}
\caption{Customised Labelling: (a) depicts a true positive AOI detection with an associated user-defined label; (b) illustrates a false positive AOI detection with a corrected user-defined label.}
\label{fig:labeling}
\end{figure}

Gaze2AOI offers capability to complement the automatically generated AOI by providing additional labels to every AOI.  
This feature is useful when YOLO prediction is inaccurate or where a more specific semantic of AOIs is required. An example is the aforementioned assignment of people's names to object classes, such as ``Human Face'' or ``Person''.
\autoref{fig:labeling} illustrates the utility of this tool. \autoref{fig:labeling} (a) shows how labelling serves as an extension to the insights gained from the YOLO results. Here, we associate the detected ``House'' with its owner. Conversely, the overlooked AOI ``Window'' does not receive any additional designation. In \autoref{fig:labeling} (b) an oven is mistakenly identified as a window. This error is rectified by assigning the new label ``Oven''. Similarly, the detected person is not assigned any additional value as they are not fully visible. Through data aggregation, AOIs lacking additional labels can be effectively filtered out.

To expedite the process, we use an algorithm to extract desired key-frames, thus avoiding the need to traverse through all frames or watch the entire video. The current version of the algorithm iterates through frames starting at the current one and selects subsequent frames containing different objects or a different number of objects. As we discuss in the future work section, there are opportunities for further enhancement in this regard.




\subsection{Optimisation}


To streamline the process, we have introduced the option to pre-run the predictions. Currently, this can be accomplished using a command-line interface (CLI) command. By running the predictions in advance, object tracking results can be generated and saved as a CSV file. Subsequently, these pre-computed predictions can be loaded concurrently with the eye-tracking data and the corresponding video. 
Pre-running predictions in advance expedites the annotation and saving of the video.
Another method to expedite the process is to omit frames from the video for which no eye-tracking data is available. In such case, the algorithm skips predictions on irrelevant frames. It is important to note that this option is at the discretion of the user, as the resulting annotated video would differ from the original.


Down sampling the number of frames of the video and the corresponding eye-tracking data is another way to speed the process. However, it is essential to consider the potential impact on the quality and accuracy of such analysis, as down sampling may lead to loss of information. Users should carefully weigh the trade-offs and determine the appropriate level of down sampling based on their specific requirements and constraints.
In the customised labelling, we also present just the relevant frames with distinct objects or object counts (to speed up the manual annotation) instead of letting the viewer watch the complete video, which would be time-consuming if there are just a few detected objects. 

\section{Conclusion and Future Work}





In this paper, we present the tool that uses a deep-learning model for object (areas of interest (AOI)) tracking as well as combining the eye-tracking data with detected AOIs. Moreover, the tool offers users the flexibility to modify the provided AOI predictions. The tool was built for processing video content after it has been already watched to post-analyse users' behaviour and attention. In order to adapt the content on the fly, real-time user behaviour analysis would be needed. Despite this, by analysing video content with our tool to better understand users the researchers and practitioners can adapt the content for future users.    

As part of our ongoing efforts to enhance the functionality and usability of our tool, several avenues for future work and improvements have been identified. One such avenue is the incorporation of unsupervised learning of face representations as in the work of Datta et al.~\cite{datta2018unsupervised} that incorporates CNN based similarity function. This approach could be particularly beneficial for distinguishing between the same human faces and avoiding redundant labelling. Additionally, exploring the potential of the new version of the YOLO model, YOLO-World~\cite{cheng2024yolo}, which can detect objects based on prompts, holds promise for simplifying user interaction by eliminating the need for check-boxes. Furthermore, enhancing the system with features such as custom drawing of bounding boxes and refining the custom labelling functionality will contribute to further improving the tool's capabilities and user experience.

\section*{Acknowledgement}
This research was funded by the Slovenian Research Agency, grant number P1-0383, P5-0433, IO-0035, J5-50155 and J7-50096. This work has also been supported by the research program CogniCom (0013103) at the University of Primorska.
\bibliographystyle{splncs04}
\bibliography{mybib}

\begin{thebibliography}{10}
\providecommand{\url}[1]{\texttt{#1}}
\providecommand{\urlprefix}{URL }
\providecommand{\doi}[1]{https://doi.org/#1}

\bibitem{barz}
Barz, M., Bhatti, O.S., Alam, H.M.T., Nguyen, D.M.H., Sonntag, D.: Interactive fixation-to-aoi mapping for mobile eye tracking data based on few-shot image classification. In: Companion Proceedings of the 28th International Conference on Intelligent User Interfaces. p. 175–178. IUI '23 Companion, Association for Computing Machinery, New York, NY, USA (2023). \doi{10.1145/3581754.3584179}, \url{https://doi.org/10.1145/3581754.3584179}

\bibitem{barz2021automatic}
Barz, M., Kapp, S., Kuhn, J., Sonntag, D.: Automatic recognition and augmentation of attended objects in real-time using eye tracking and a head-mounted display. In: ACM Symposium on Eye Tracking Research and Applications. pp.~1--4 (2021)

\bibitem{bell1823xv}
Bell, C.: Xv. on the motions of the eye, in illustration of the uses of the muscles and nerves of the orbit. Philosophical Transactions of the Royal Society of London (113),  166--186 (1823)

\bibitem{blascheck2014state}
Blascheck, T., Kurzhals, K., Raschke, M., Burch, M., Weiskopf, D., Ertl, T.: State-of-the-art of visualization for eye tracking data. In: Eurovis (stars). p.~29 (2014)

\bibitem{carter2020best}
Carter, B.T., Luke, S.G.: Best practices in eye tracking research. International Journal of Psychophysiology  \textbf{155},  49--62 (2020)

\bibitem{cheng2024yolo}
Cheng, T., Song, L., Ge, Y., Liu, W., Wang, X., Shan, Y.: Yolo-world: Real-time open-vocabulary object detection. arXiv preprint arXiv:2401.17270  (2024)

\bibitem{datta2018unsupervised}
Datta, S., Sharma, G., Jawahar, C.: Unsupervised learning of face representations. In: 2018 13th IEEE International Conference on Automatic Face \& Gesture Recognition (FG 2018). pp. 135--142. IEEE (2018)

\bibitem{beugher}
De~Beugher, S., Ichiche, Y., Br\^{o}ne, G., Goedem\'{e}, T.: Automatic analysis of eye-tracking data using object detection algorithms. In: Proceedings of the 2012 ACM Conference on Ubiquitous Computing. p. 677–680. UbiComp '12, Association for Computing Machinery, New York, NY, USA (2012). \doi{10.1145/2370216.2370363}, \url{https://doi.org/10.1145/2370216.2370363}

\bibitem{ducasse2022interactive}
Ducasse, J., Kljun, M., Attygalle, N.T., Pucihar, K.{\v{C}}.: Interactive web documentaries: A case study of video viewing behaviour on iotok. International Journal of Human--Computer Interaction  \textbf{38}(10),  949--972 (2022)

\bibitem{julie}
Ducasse, J., Kljun, M., {\v{C}}opi{\v{c}}~Pucihar, K.: Interactive web documentaries: A case study of audience reception and user engagement on iotok. International Journal of Human--Computer Interaction  \textbf{36}(16),  1558--1584 (2020)

\bibitem{duchowski}
Duchowski, A.T., Gehrer, N.A., Sch\"{o}nenberg, M., Krejtz, K.: Art facing science: Artistic heuristics for face detection: tracking gaze when looking at faces. In: Proceedings of the 11th ACM Symposium on Eye Tracking Research \& Applications. ETRA '19, Association for Computing Machinery, New York, NY, USA (2019). \doi{10.1145/3317958.3319809}, \url{https://doi.org/10.1145/3317958.3319809}

\bibitem{fehringer}
Fehringer, B.C.O.F.: Eye tracking gaze visualiser: eye tracker and experimental software independent visualisation of gaze data. In: Proceedings of the Symposium on Eye Tracking Research and Applications. p. 259–262. ETRA '14, Association for Computing Machinery, New York, NY, USA (2014). \doi{10.1145/2578153.2578191}, \url{https://doi.org/10.1145/2578153.2578191}

\bibitem{OpenImages}
Kuznetsova, A., Rom, H., Alldrin, N., Uijlings, J., Krasin, I., Pont-Tuset, J., Kamali, S., Popov, S., Malloci, M., Kolesnikov, A., Duerig, T., Ferrari, V.: The open images dataset v4: Unified image classification, object detection, and visual relationship detection at scale. IJCV  (2020)

\bibitem{storageOpenImages}
{Open Images}: {Open Images V7 --- storage.googleapis.com}. \url{https://storage.googleapis.com/openimages/web/index.html} (2022)

\bibitem{redmon2016you}
Redmon, J., Divvala, S., Girshick, R., Farhadi, A.: You only look once: Unified, real-time object detection. In: Proceedings of the IEEE conference on computer vision and pattern recognition. pp. 779--788 (2016)

\bibitem{ultralyticsHome}
Ultralytics: {H}ome --- docs.ultralytics.com. \url{https://docs.ultralytics.com/}, [Accessed 15-04-2024]

\bibitem{wadeMoving}
Wade, N., Tatler, B.W.: The moving tablet of the eye: The origins of modern eye movement research. Oxford University Press (2005)

\bibitem{wolf2018automating}
Wolf, J., Hess, S., Bachmann, D., Lohmeyer, Q., Meboldt, M.: Automating areas of interest analysis in mobile eye tracking experiments based on machine learning. Journal of Eye Movement Research  \textbf{11}(6) (2018)

\bibitem{xiao2020review}
Xiao, Y., Tian, Z., Yu, J., Zhang, Y., Liu, S., Du, S., Lan, X.: A review of object detection based on deep learning. Multimedia Tools and Applications  \textbf{79},  23729--23791 (2020)

\bibitem{zhang2018complete}
Zhang, X., Yuan, S.M., Chen, M.D., Liu, X.: A complete system for analysis of video lecture based on eye tracking. Ieee Access  \textbf{6},  49056--49066 (2018)

\end{thebibliography}
%






\end{document}